\documentclass{jetpl}
\usepackage {cite}
\twocolumn

\lat

\title{Effect of long-range interactions on the Kohn-Luttinger
mechanism of the Cooper instability in the Shubin-Vonsowsky model}

\rtitle{Effect of long-range interactions on the Kohn-Luttinger
mechanism of the Cooper instability\ldots}

\sodtitle{Effect of long-range interactions on the Kohn-Luttinger
mechanism of the Cooper instability in the Shubin-Vonsowsky model}

\author{M.\,Yu.\,Kagan$^{+}$,
V.\,V.\,Val'kov$^{*,\bigtriangleup}$, V.\,A.
Mitskan$^{*,\bigtriangleup}$, M.\,M.
Korovushkin$^{*,\bigtriangleup}$}

\rauthor{M.\,Yu.\,Kagan V.\,V.\,Val'kov, V.\,A. Mitskan, M.\,M.
Korovushkin}

\sodauthor{}

\address{$^+$P.\,L. Kapitza Institute for Physical Problems, 119334 Moscow, Russia\\~\\
$^*$L.\,V. Kirensky Institute of Physics, 660036 Krasnoyarsk,
Russia\\~\\
$^\bigtriangleup$Siberian State Aerospace University, 660014
Krasnoyarsk, Russia}

\dates{24 January 2013}{*}

\abstract{The problem of Cooper instability is studied by
implementing the Kohn-Luttinger mechanism in the framework of the
Shubin-Vonsowsky model taking into account the intersite Coulomb
interactions within the first and second coordination spheres. It
is shown that the Kohn-Luttinger renormalization for the effective
interaction in the second-order terms of perturbation theory, as
well as the inclusion of intersite hoppings in the second and
third coordination spheres, significantly affects the interplay
between the superconducting phases with $d_{xy}-$, $p-$, $s-$, and
$d_{x^2-y^2}-$wave symmetries of the order parameter.}


\begin{document}

\maketitle

\section{Introduction}

The Kohn-Luttinger mechanism~\cite{Kohn65} attracts considerable
current interest as a promising non-phonon mechanism of
superconductivity applicable for many physical systems such as
cuprate superconductors~\cite{Kagan88} and doped
graphene~\cite{Gonzalez08,Nandkishore12}, as well as the mechanism
underlying superfluidity in $^3$He~\cite{Kagan94, Volhard90,
Volovik03} and in topological superfluids~\cite{Marienko12}. The
$d_{x^2-y^2}-$wave superconductivity arising in cuprates
stimulated intense studies of the interplay between
superconducting phases with different types of symmetry in the
framework of the Hubbard model in the $U\ll W$
limit~\cite{Kagan91,Baranov92,Hlubina99,Raghu10,Scalapino86,Zanchi96}.

The effect of the screened Coulomb interaction on the
Kohn-Luttinger mechanism for cuprate superconductors was recently
studied in~\cite{Alexandrov11}. A renewed interest in the effect
of the long-range Coulomb correlations on the structure of the
phase diagrams for the Mott-Hubbard systems at low electron
density $n$ led to the additional studies~\cite{Kagan11,Raghu12}
based on the Shubin--Vonsowsky model~\cite{Shubin34,Shubin35},
which takes into account the interactions ($V$) between electrons
located at different lattice sites. In~\cite{Raghu12}, the phase
diagram in the $n-V$ plane was constructed for such a model. This
phase diagram demonstrates the result of the interplay between
superconducting phases with different types of symmetry. The
corresponding calculations involve only the intersite hopping
within the first coordination sphere and the intersite Coulomb
interactions are considered including only the first-order terms
of perturbation theory. Since the polarization Kohn--Luttinger
contributions manifest themselves only in the second order, the
analysis of the effects related to such contributions appears to
be promising for finding out the ranges of existence for
superconducting phases with different symmetries.

In this paper, the Cooper instability for the Shubin-–Vonsowsky
model is studied in the weak coupling limit of the Born
approximation ($W > U > V$) taking into account the long-range
hopping processes and the intersite Coulomb interactions within
the first and second coordination spheres. In the calculation of
the effective interaction in the Cooper channel, we include the
polarization contributions, graphically represented by four
Kohn--Luttinger diagrams (Fig.~\ref{diagrams}). We demonstrate
that the long-range Coulomb interactions and the long-range
intersite electron hopping produce a pronounced effect on the
conditions needed for the Cooper pairing with $s-$, $p-$, and
$d-$wave symmetries of the order parameter. In particular, they
illustrate the possibility of the $d_{x^2-y^2}-$type of pairing
arising. Note that, in addition to the Shubin--Vonsowsky model,
the $t-J$ model also remains a highly probable challenger for the
adequate description of high-$T_c$ superconductivity with $d-$wave
pairing~\cite{Kagan_Rice94,Plakida03_01,Izumov99,Belinicher97_95}.
A consideration of the Coulomb interaction within the framework of
the $t-J$ model will be a subject of our further studies.

\section{Model}

The Hamiltonian of the Shubin--Vonsowsky model in the
quasimomentum representation has the form
\begin{eqnarray}
\hat{H}
&=&\sum\limits_{\textbf{p}\sigma}(\varepsilon_{\textbf{p}}-\mu)
c^{\dagger}_{\textbf{p}\sigma}c_{\textbf{p}\sigma} +
U\sum_{\textbf{p}\textbf{p'}\textbf{q}}c^{\dagger}_{\textbf{p}\uparrow}
c^{\dagger}_{\textbf{p'}+\textbf{q}{\downarrow}}c_{\textbf{p}+\textbf{q}{\downarrow}}
c_{\textbf{p'}{\uparrow}}\nonumber\\
&+&\sum_{\textbf{p}\textbf{p'}\textbf{q}\sigma\sigma'}V_{\textbf{p}-\textbf{p'}}\,c^{\dagger}_{\textbf{p}\sigma}
c^{\dagger}_{\textbf{p'}+\textbf{q}{\sigma'}}c_{\textbf{p}+\textbf{q}{\sigma'}}c_{\textbf{p'}{\sigma}},
\end{eqnarray}
where the electron energy including the long-range intersite
hopping with the intensity determined by the parameters $t_2$ and
$t_3$ is described by the expression
\begin{eqnarray}
\varepsilon_{\textbf{p}}&=&2t_1(\textrm{cos}\,p_x+\textrm{cos}\,p_y)+
4t_2\textrm{cos}\,p_x\textrm{cos}\,p_y \nonumber\\
&+&2t_3(\textrm{cos}\,2p_x+\textrm{cos}\,2p_y).
\end{eqnarray}
The Fourier transform corresponding to the Coulomb repulsion
between electrons located at the nearest-neighbor or
next-nearest-neighbor sites can be written as
\begin{eqnarray}
V_{\textbf{p}}=2V_1(\textrm{cos}\,p_x+\textrm{cos}\,p_y)+4V_2
\textrm{cos}\,p_x\textrm{cos}\,p_y.
\end{eqnarray}
The utilization of the aforementioned Born approximation in the
weak-coupling limit allows us to use only the diagrams
corresponding to the first and second orders of perturbation
theory in terms of the coupling constant.

The opposite (strong-coupling) limit, $U > V > W$, was studied
in~\cite{Kagan11}. In this case, the inclusion of the first- and
second-order diagrams is justified only in the low electron
density limit ($n\ll 1$), for which the Fermi gas type
Galitskii-Bloom expansion~\cite{Galitskii58,Bloom75} works well.
Only the main exponential term for $T_c$ is calculated in this
work, as well as in~\cite{Kagan11}. The accurate evaluation of the
pre-exponential factor requires the inclusion of the third and
fourth-order diagrams. The Born approximation used in this paper
allows us to consider also higher values of the electron density.

It is well known that the intersite Coulomb interaction suppresses
the Cooper pairing in the first order of perturbation theory. The
contributions to the effective interaction that improve the
conditions favoring the Cooper instability appear in the second
order. It is also important that the inclusion of the distant
hopping provides an opportunity to shift the position of the Van
Hove singularity in the electron density of states toward lower
densities. In this paper, we analyze only the range of electron
densities for which we do not approach too close to the Van Hove
singularities, in order to avoid the summation of parquet
diagrams~\cite{Dzyaloshinskii88}. In the Mott-Hubbard systems, the
screening radius can exceed the unit cell size~\cite{Zaitsev-04}.
This determines the efficiency of the Shubin--Vonsowsky model,
which takes into account the intersite Coulomb interaction within
several coordination spheres. In this case, the effects related to
the Brillouin zone manifest themselves in the momentum dependence
of $V_\textbf{p}$, which is described by a periodic function, and
are clearly pronounced.

\section{RENORMALIZED INTERACTION
IN THE COOPER CHANNEL}

\begin{figure}[t!]
\begin{center}
\includegraphics[width=0.4\textwidth]{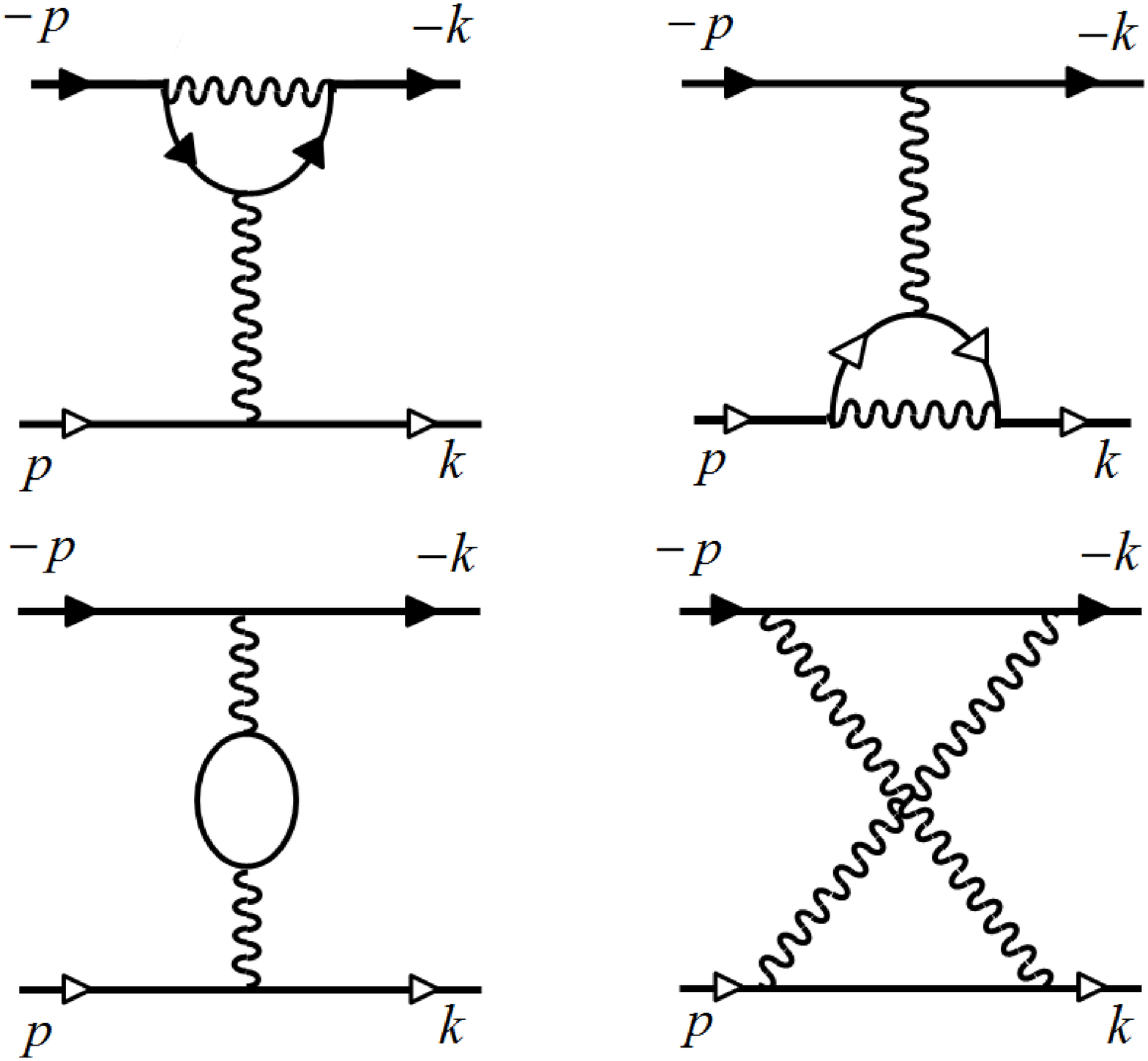}
\caption{Fig. 1. Second-order diagrams for the effective
interaction.} \label{diagrams}
\end{center}
\end{figure}

The second-order correction $\delta\tilde{\Gamma}(p, k)$ for the
effective interaction in the Cooper channel is determined by four
Kohn-Luttinger diagrams shown in Fig.~\ref{diagrams}. Solid lines
with the light (dark) arrows correspond to the Green's functions
of electrons with spin projections equal to +1/2 (–1/2). In these
diagrams, the existence of two solid lines without arrows implies
the performed summation over the values of the spin projections.
The wavy lines correspond to the unrenormalized interactions. The
scattering of electrons with the same spin projection gives rise
only to the intersite contribution. If we have the interaction
between electrons with opposite spins, the scattering amplitude is
determined by the sum of the Hubbard and intersite interactions.
Therefore, when we deal only with the Hubbard repulsion, the
$\delta\tilde{\Gamma}(p, k)$ correction for the effective
interaction is given only by the fourth diagram. If we take into
account the Coulomb repulsion at the neighboring sites, all four
diagrams shown in Fig.~\ref{diagrams} contribute to the
renormalized amplitude.

Introducing analytical expressions for the diagrams in
Fig.~\ref{diagrams} (by using four-vectors
$p\equiv(\textbf{p},i\omega_n)$ and
$k\equiv(\textbf{k},i\omega_m)$) and performing the summation over
the Matsubara frequencies, we arrive at the expression for the
effective interaction
\begin{eqnarray}\label{Gamma_wave}
&&\widetilde{\Gamma}(p,k)=U+V_{\textbf{p}-\textbf{k}}+
\delta\widetilde{\Gamma}(p,k),
\end{eqnarray}
\begin{eqnarray}
&&\delta\widetilde{\Gamma}(p,k)=\frac1N\sum_{\textbf{p}_1}(U+V_{\textbf{p}-\textbf{k}})
(2V_{\textbf{p}-\textbf{k}}-V_{\textbf{p}_1+\textbf{p}}-V_{\textbf{p}_1-\textbf{k}})\nonumber\\
&&\qquad\qquad\times\frac{f(\varepsilon_{\textbf{p}_1})-f(\varepsilon_{\textbf{p}_1+\textbf{p}-\textbf{k}})}
{i\omega_n-i\omega_m+\varepsilon_{\textbf{p}_1}-\varepsilon_{\textbf{p}_1+\textbf{p}-\textbf{k}}}\nonumber\\
&&\qquad\qquad+\frac1N\sum_{\textbf{p}_1}(U+V_{\textbf{p}_1-\textbf{p}})(U+V_{\textbf{p}_1-\textbf{k}})\nonumber\\
&&\qquad\qquad\times\frac{f(\varepsilon_{\textbf{p}_1})-f(\varepsilon_{\textbf{p}_1-\textbf{p}-\textbf{k}})}
{i\omega_n+i\omega_m-\varepsilon_{\textbf{p}_1}+
\varepsilon_{\textbf{p}_1-\textbf{p}-\textbf{k}}},\nonumber
\end{eqnarray}
where $f(\varepsilon)=(\exp(\frac{\varepsilon-\mu}{T})+1)^{-1}$.

\section{BETHE-SALPETER EQUATION}

\begin{figure}[t!]
\begin{center}
\includegraphics[width=0.5\textwidth]{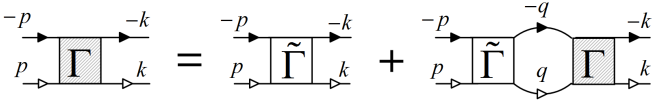}
\caption{Fig. 2. Bethe--Salpeter equation.} \label{bethe_eq}
\end{center}
\end{figure}

Utilizing the renormalized interaction in the weak-coupling
approximation, we get the ladder-type equation for the scattering
amplitude in the Cooper channel (see Fig.~\ref{bethe_eq}). In the
explicit form, this equation reads as
\begin{eqnarray}
\label{BetheEq}
&&\Gamma(p\,|k)=\widetilde{\Gamma}(p\,|k)\\
&&-\frac
TN\sum_{\textbf{q},\omega_l}\widetilde{\Gamma}(p\,|\textbf{q},i\omega_l)
\frac{\Gamma(\textbf{q},i\omega_l|k)}{(i\omega_l-\xi_{\textbf{q}})(-i\omega_l-\xi_{\textbf{q}})}
,\nonumber
\end{eqnarray}
where $\xi_{\textbf{q}}=\varepsilon_{\textbf{q}}-\mu$. Here,
performing the summation over the Matsubara frequencies
$\omega_l$, we take into account that the main contribution to the
sum in the Cooper channel in the weak-coupling approximation comes
from the $\omega_l$ values close to zero. Therefore, we can
neglect the frequency dependence of the renormalized interaction.
In this case, the total amplitude is independent of the frequency
and the summation over $\omega_l$ in the Bethe--Salpeter equation
can be performed explicitly. As a result, we get the following
integral equation, which determines the momentum dependence of the
scattering amplitude
\begin{eqnarray}
\label{IntegralEq} \Gamma(\textbf{p}\,|\textbf{k})=
\widetilde{\Gamma}(\textbf{p}\,|\textbf{k})-\frac
1N\sum_{\textbf{q}}\widetilde{\Gamma}(\textbf{p}\,|\textbf{q})
\frac{\displaystyle\tanh\biggl(\frac{\xi_{\textbf{q}}}{2T}\biggr)}{2\xi_{\textbf{q}}}
\Gamma(\textbf{q}\,|\textbf{k}).
\end{eqnarray}
It is well known that the pole corresponding to the Cooper
instability can be found analyzing the homogeneous part of the
reduced equation~\cite{Gorkov61}. Introducing the integration over
the isoenergetic curves and taking into account that the main
contribution comes from the contours close to the Fermi
contour~\cite{Baranov92,Hlubina99,Raghu10,Alexandrov11}, we get
the equation
\begin{equation}
\label{IntegralEqPhi}
\frac{1}{(2\pi)^2}\oint\limits_{\varepsilon_{\textbf{q}}=\mu}
\frac{d\hat{\textbf{q}}} {v_F(\hat{\textbf{q}})}
\widetilde{\Gamma}(\hat{\textbf{p}}\,|\hat{\textbf{q}})
\Gamma(\hat{\textbf{q}})=\lambda\Gamma(\hat{\textbf{p}}),
\end{equation}
where $\lambda^{-1}\simeq \ln(T_c/W)$, quasimomenta
$\hat{\textbf{p}}$ and $\hat{\textbf{q}}$ lie on the Fermi
surface, $v_F(\hat{\textbf{q}})$ is the Fermi velocity. To solve
this equation, we should consider an eigenvalue problem.

We represent the kernel of integral equation (\ref{IntegralEqPhi})
as a superposition of the functions, each belonging to one of the
irreducible representations of the $C_{4v}$ symmetry group on the
square lattice. It is well known that this group has five
irreducible representations~\cite{Landau89}. For each
representation, Eq.~(\ref{IntegralEqPhi}) has a solution with its
own effective coupling constant $\lambda$. Further on, we use the
following notation to classify the symmetries of the order
parameter: representations $A_1$, $A_2$, $B_1$, $B_2$, and $E$
correspond to the $s-$wave, extended $s-$wave, $d_{xy}-$wave,
$d_{x^2-y^2}-$wave, and $p-$wave types of symmetry, respectively.

We seek a solution of Eq.~(\ref{IntegralEqPhi}) in the form
\begin{equation}\label{solution}
\Gamma(\phi)=\sum\limits_{\alpha\,n}\Delta_{\alpha\,n}g_{\alpha\,n}(\phi),
\end{equation}
where $\alpha$ is the ordinal number of the representation, $n$ is
the ordinal number of a function belonging to the given
representation, and $\phi$ is the angle characterizing the
direction of the quasimomentum $\hat{\textbf{p}}$ with respect to
the $p_x$ axis. The explicit form of the orthonormalized functions
$g_{\alpha\,n}(\phi)$ is given by the expressions
\begin{eqnarray}\label{harmon}
&&A_1\rightarrow~g_{s,n}(\phi)=\frac{1}{\sqrt{(1+\delta_{n0})\pi}}\,
\textrm{cos}\,4n\phi,~~n\in[\,0,\infty),\label{invariants_s}\nonumber\\
&&A_2\rightarrow~g_{s_{ext},n}(\phi)=\frac{1}{\sqrt{\pi}}\,\textrm{sin}\,
4(n+1)\phi,\label{invariants_s1}\nonumber\\
&&B_1\rightarrow~g_{d_{xy},n}(\phi)=\frac{1}{\sqrt{\pi}}\,
\textrm{sin}\,(4n+2)\phi,\label{invariants_dxy}\\
&&B_2\rightarrow~g_{d_{x^2-y^2},n}(\phi)=\frac{1}{\sqrt{\pi}}\,
\textrm{cos}\,(4n+2)\phi,\label{invariants_dx2y2}\nonumber\\
&&E~\rightarrow~g_{p,n}(\phi)=\frac{1}{\sqrt{\pi}}\,(A\,\textrm{sin}\,
(2n+1)\phi+B\,\textrm{cos}\,(2n+1)\phi)\label{invariants_p}\nonumber.
\end{eqnarray}

Substituting Eq.~(\ref{solution}) into Eq.~(\ref{IntegralEqPhi}),
performing integration over the angles, and using the
orthonormality condition for the functions $g_{\alpha\,n}(\phi)$,
we find
\begin{equation}\label{EqDelta}
\hat{\Lambda}_{\alpha n;\beta m}\Delta_{\beta
m}=\lambda\Delta_{\alpha n},
\end{equation}
where
\begin{eqnarray}
\label{matrix} \hat{\Lambda}_{\alpha n;\beta
m}&=&\frac{1}{(2\pi)^2}\int\limits_0^{2\pi}d\phi_{\textbf{p}}
\int\limits_0^{2\pi}d\phi_{\textbf{q}}\frac{d\hat{\textbf{q}}}
{d\phi_{\textbf{q}}v_F(\hat{\textbf{q}})}
\widetilde{\Gamma}(\hat{\textbf{p}}\,|\hat{\textbf{q}})\nonumber\\
&\times&g_{\alpha n}(\phi_{\textbf{p}}) g_{\beta
m}(\phi_{\textbf{q}}).
\end{eqnarray}
Since $T_c\sim\exp\bigl(1/\lambda \bigr)$, each negative
eigenvalue $\lambda$ corresponds to a superconducting phase with
the specified symmetry type of the order parameter. Each solution
corresponds to only one irreducible representation, but its
expansion in terms of the basis functions generally includes
several angular harmonics. The highest critical temperature
corresponds to the largest absolute value of $\lambda$.
\begin{figure}[t!]
\begin{center}
\includegraphics[width=.4\textwidth]{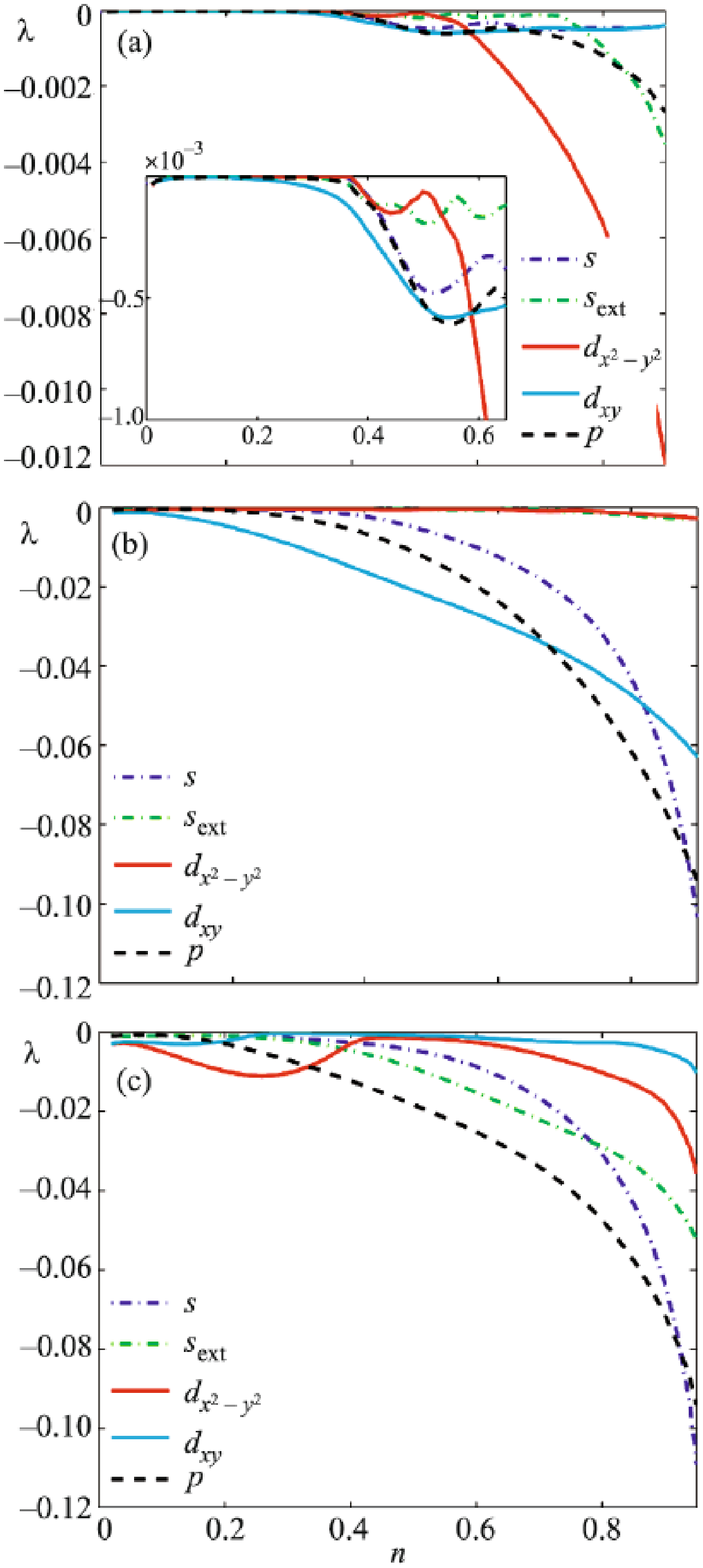}
\caption{Fig. 3. Electron density dependence of $\lambda$ for
$t_2=t_3=0$ and $U=1$ at different values of the parameters
characterizing the intersite Coulomb interaction: (a)
$V_1=V_2=0$;\quad (b) $V_1=0.5,~V_2=0$;\quad (c)
$V_1=0.5,~V_2=0.25$.} \label{Tc_V0}
\end{center}
\end{figure}

\section{RESULTS AND DISCUSSION}

In Fig.~\ref{Tc_V0}, we illustrate the dependence of the effective
coupling constant $\lambda$ on the electron density $n$ for
different types of symmetry of the superconducting order
parameter. The calculations were performed at $t_2 = t_3 = 0$ and
$U = 1$ (all energy parameters are measured in units of $|t_1|$)
at different values of the parameters $V_1$ and $V_2$, which
characterize the intersite Coulomb interactions. In
Fig.~\ref{Tc_V0} (a), we plot the $\lambda(n)$ curve in the
absence of the intersite Coulomb interaction ($V_1 = 0$ and $V_2 =
0$). It agrees well with the corresponding curves reported
in~\cite{Hlubina99}. At low electron densities ($n = 0 - 0.52$),
in the first two orders of perturbation theory, we get
superconductivity with the $d_{xy}-$wave type of the order
parameter~\cite{Baranov92}. In the range $n = 0.52 - 0.58$, the
ground state corresponds to the phase with the $p-$wave order
parameter. At $n>0.58$, the $d_{x^2-y^2}-$wave type of
superconductivity is the dominant one.
\begin{figure}[t!]
\begin{center}
\includegraphics[width=.4\textwidth]{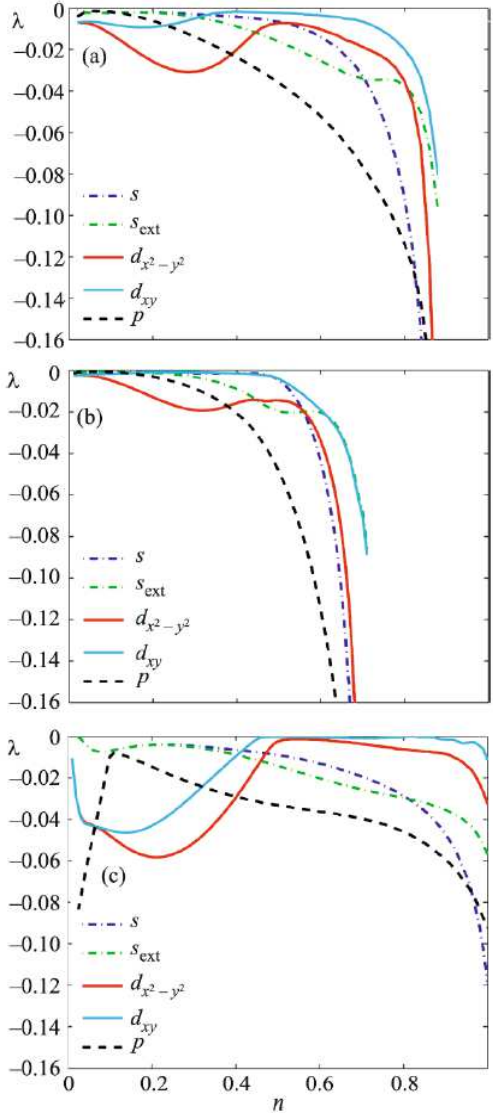}
\caption{Fig. 4. Electron density dependence of $\lambda$ for
$U=1$; $V_1=0.5,~V_2=0.25$ at different values of the distant
hopping integrals: (a) $t_2=0.15,~t_3=0$;\quad (b)
$t_2=0.15,~t_3=-0.1$;\quad (c) $t_2=0.15,~t_3=0.1$.}\label{Tc_VV}
\end{center}
\end{figure}

\begin{figure}[h!]
\begin{center}
\includegraphics[width=.45\textwidth]{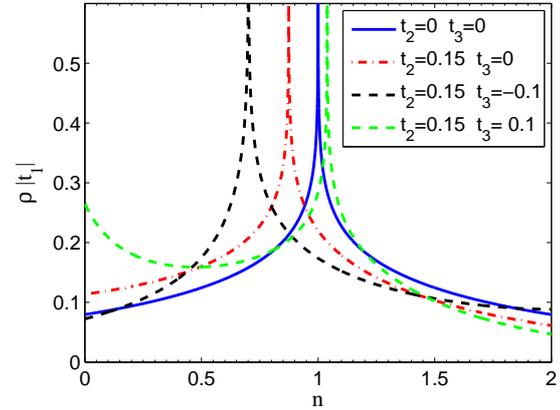}
\caption{Fig. 5. Evolution of the electron density of states under
variation of the hopping integrals.} \label{DOS}
\end{center}
\end{figure}

The inclusion of the intersite Coulomb interaction significantly
affects the interplay between the superconducting phases. This is
clearly seen in Fig.~\ref{Tc_V0} (b), where we represent the
$\lambda(n)$ plots for $V_1 = 0.5$ and $V_2 = 0$. Their comparison
to the plots in Fig.~\ref{Tc_V0} (a) demonstrates that the range
of dominance for the $d_{xy}-$ and $p-$wave phases becomes
broader. At high electron densities ($n\geq0.89$), we have the
$s-$wave type of pairing, for which the main contribution comes
from the $g_{s,1}(\phi)=\displaystyle\frac{1}{\sqrt{\pi}}\cos
4\phi$ harmonics.

It is necessary to note that the Cooper pairing calculated up to
the second order of perturbation theory in terms of the Coulomb
interaction is not suppressed by the first-order terms. This stems
from the fact that the bare Coulomb repulsion
$V_{\textbf{p}-\textbf{k}}$ suppresses only a part of the
harmonics, whereas the other harmonics lead to the Cooper
instability. For example, in the range of existence for the
$p-$wave pairing, the main contribution of the bare repulsion
$V_{\textbf{p}-\textbf{k}}$ in the $p-$wave channel is given by
the function
$g_{p,0}(\phi)=\displaystyle\frac{1}{\sqrt{\pi}}\,(A\,\textrm{sin}\,
\phi+B\,\textrm{cos}\,\phi)$, whereas the main contribution to the
scattering amplitude in the $p-$channel comes from the function
$g_{p,1}(\phi)=\displaystyle\frac{1}{\sqrt{\pi}}\,(A\,\textrm{sin}\,
3\phi+B\,\textrm{cos}\,3\phi)$.

In Fig.~\ref{Tc_V0} (c), we represent the $\lambda(n)$ curves
plotted taking into account the intersite Coulomb interactions
within the first and second coordination spheres ($V_1 = 0.5$ and
$V_2 = 0.25$). The comparison with Fig.~\ref{Tc_V0} (b) suggests
that the inclusion of the long-range Coulomb repulsion $V_2$
favors the $d_{x^2-y^2}-$wave type of pairing at low charge
carrier densities ($n = 0.05-0.34$).

Distant electron hoppings (to the sites located outside the first
coordination sphere) significantly affect the interplay of
different superconducting phases. This is illustrated in
Fig.~\ref{Tc_VV}, where we show the $\lambda(n)$ curves plotted at
$U = 1,\,V_1 = 0.5$, and $V_2 = 0.25$ for different values of
$t_2$ and $t_3$. The plots shown in Fig.~\ref{Tc_VV} (a) are
calculated with the inclusion of the electron hoppings within the
first two coordination spheres ($t_2 = 0.15,\,t_3 = 0$). At these
parameters, the critical density of fermions $n_{vH}$
(corresponding to the Van Hove singularity) shifts from the half
filling toward lower electron densities (Fig.~\ref{DOS}).
Comparing Figs.~\ref{Tc_V0} (c) and ~\ref{Tc_VV} (a), we see that
the inclusion of the distant hoppings $t_2$ leads to broadening of
the dominance range for the $d_{x^2-y^2}-$type of pairing to $n =
0.4$ and to an increase in the absolute value of $\lambda$ in this
region.

Figures~\ref{Tc_VV} (b) and \ref{Tc_VV} (c) show the $\lambda(n)$
dependence calculated with the additional inclusion of hoppings to
the third coordination sphere with $t_3 < 0$ and $t_3 > 0$,
respectively. Comparison of these plots indicates that the
inclusion of hoppings with $t_3 > 0$ leads to the additional
broadening of the dominance range for the $d_{x^2-y^2}-$wave type
of pairing at low charge carrier densities and to the enhancement
of the effective interaction in this region.

\begin{figure}[t!]
\begin{center}
\includegraphics[width=.4\textwidth]{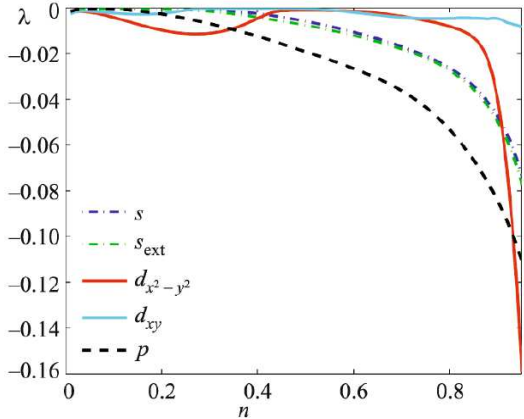}
\caption{Fig. 6. Carrier density dependence of $\lambda(n)$ at
$t_2=t_3=0$, $V_1=0.5,~V_2=0.25$ and $U=2$.} \label{Tc_U2}
\end{center}
\end{figure}

Note that with the growth of $U$, the superconducting phase with
the $d_{x^2-y^2}-$wave symmetry of the order parameter becomes
dominant in the density range close to the Van Hove singularities.
This can be seen in Fig.~\ref{Tc_U2}, where we demonstrate the
$\lambda(n)$ curves calculated at $U = 2$. In the regions of low
and high densities, the $d_{x^2-y^2}$ phase corresponds to the
ground state of the system. This result seems to be important for
the analysis of the mechanisms underlying the high-$T_c$
superconductivity. Note in this context that the critical
temperatures $T_c \sim 100\, K$ appear at $U = 3$. However, this
case is on the verge of the applicability range of the
weak-coupling approximation, which we use in this paper.

\section{Conclusions}

The analysis of the Cooper instability performed within the
framework of the Shubin--Vonsowsky model in the weak-coupling
limit ($W > U > V$) demonstrates that the Kohn--Luttinger
contributions of the polarization type calculated up to the second
order of perturbation theory lead to a significant renormalization
of the effective interaction. Note that the inclusion of the
additional Coulomb repulsion at the neighboring lattice sites
($V_1$) and even at the next-nearest-neighbor sites ($V_2$) cannot
suppress all superconducting channels in the calculations of the
effective interaction $\tilde{\Gamma}$ in the first two orders of
perturbation theory.

At $V_1\neq0$ and $V_2\neq0$ superconductivity arises in the $p-$,
$d_{xy}-$, and  $d_{x^2-y^2}-$wave channels at low and
intermediate electron densities. In the $d_{x^2-y^2}-$wave and in
the specific $s-$wave channel ($\cos{4\phi}$ harmonics), it arises
near the Van Hove singularity at high electron densities. Near the
Van Hove singularity, with the growth of the Hubbard repulsion
$U$, we get the usual $d_{x^2-y^2}-$wave type of pairing typical
of cuprates with rather realistic values of the superconducting
transition temperature. The inclusion of the distant hoppings
$t_2\neq0$ and $t_3\neq0$ shifts the Van Hove singularity toward
lower electron densities but does not change the gross phase
diagram describing the superconducting states.

We are grateful to A.\,S. Alexandrov, D.\,V. Efremov, V.\,V.
Kabanov, Yu.\,V. Kopaev, K.\,I. Kugel', M.\,S. Mar'enko, N.\,M.
Plakida, and A.\,V. Chubukov for numerous fruitful discussions and
permanent interest in this work. This work was supported by the
Presidium of the Russian Academy of Sciences (program no. 20.7
"Quantum Mesoscopic and Disordered Structures"), by the Russian
Foundation for Basic Research (project nos. 11-02-00741 and
12-02-31130), and by the Ministry of Education and Science of the
Russian Federation (state contract no. 16.740.11.0644, federal
program "Human Capital for Science and Education in Innovative
Russia" for 2009–2013). M.\,M.\,K. acknowledges the support of the
Council of the President of the Russian Federation for Support of
Young Scientists and Leading Scientific Schools (project no.
MK-526.2013.2) and the Dynasty Foundation.

\end{document}